\documentclass[12pt]{article}
\newcommand{\dagg}{{\scriptscriptstyle\dagger}}

\begin{document}
%%%%%%%%%%%%%%%%%%%%%%%%%%%%%%%
\begin{center}
{\bf
A note on ferromagnetism in the Hubbard model on the complete graph
}
\par\bigskip
Andreas Mielke\footnote{
Institut f{\"u}r theoretische Physik, Universit{\"a}t Heidelberg,
Philosophenweg 19, 69120 Heidelberg, Germany,
mielke@hybrid.tphys.uni-heidelberg.de
}
and
Hal Tasaki\footnote{
Department of Physics, Gakushuin University,
Mejiro, Toshima-ku, Tokyo 171, Japan,
hal.tasaki@gakushuin.ac.jp
}
\end{center}
%%%%%%%%%%%%%%%%%%%%%%%%%%%%%%%
\begin{abstract}
Recently there have appeared some papers which discuss the existence
of ferromagnetism in the Hubbard model defined on the complete
graph.
At least for the special electron number $N_{\rm e}=N-1$, where $N$
denotes the number of sites in the lattice, the existence of
ferromagnetism in this model was established rigorously some time ago,
 as
special (and indeed the simplest) cases of more general classes of
models.
Here we explain these implications
to clarify the situation,
although we believe the implications are straightforward.

We are posting this note to the preprint archive to make it public,
but we are not
planning to publish it in other forms.
This is because we do not think the problem warrants any extra
publications, and we believe that
the validity of our explanation is evident to the readers.
\end{abstract}
%%%%%%%%%%%%%%%%%%%%%%%%%%%%%%%
\section{The problem}
\label{pr}
Since our motivation is explained in the abstract,
we start by defining
the problem precisely.
Let $N$ be a positive integer.
We identify our lattice $\Lambda$ with the set of integers
$\{1,2,\ldots,N\}$.
For $i\in\Lambda$ and $\sigma=\uparrow,\downarrow$,
we let $c^{\dagg}_{i,\sigma}$, $c_{i,\sigma}$, and
$n_{i,\sigma} = c^{\dagg}_{i,\sigma}c_{i,\sigma}$
the standard creation, annihilation, and number operators,
respectively, for an electron
at site $i$ with spin $\sigma$.
The model under consideration has the Hamiltonian
\begin{equation}
	H = t\sum_{\sigma=\uparrow,\downarrow}
	\sum_{i,j\in\Lambda;i\ne j}c^{\dagg}_{i,\sigma}c_{j,\sigma}
	+U\sum_{i\in\Lambda}n_{i,\uparrow}n_{i,\downarrow},
	\label{Ham}
\end{equation}
where the hopping amplitude satisfies $t>0$ and the
Coulomb interaction $U>0$.
Another important parameter of the model is
the electron number $N_{\rm e}$,
which is the eigenvalue of
$\sum_{\sigma=\uparrow,\downarrow}
\sum_{i\in\Lambda}(n_{i,\uparrow}+n_{i,\downarrow})$.

Among the statements discussed in \cite{Salerno,Wang}
is the following.
\par\bigskip\noindent
{\bf Corollary:}
In the Hilbert space where  the electron number is fixed to
$N_{\rm e}=N-1$, the ground states of (\ref{Ham}) exhibit
saturated ferromagnetism
and are nondegenerate\footnote{
When $N_{\rm e}<N-1$, one can easily construct
ferromagnetic and non-ferromagnetic
ground states.
} apart from the trivial spin degeneracy for any
$t>0$ and $U>0$.
\par\bigskip
As we mentioned in the abstract, the above Corollary follows as
special (and the easiest) cases of general results in
\cite{Andreas} and in \cite{AndreasHal}.
It is also straightforward to get the Corollary
from Nagaoka's theorem \cite{Nagaoka}.
We discuss these three proofs briefly\footnote{
The variational argument of \cite{Wang} does not give a proof.
}.
See \cite{Hal,later} for the progress of the research in ferromagnetism
which followed \cite{Andreas,AndreasHal}.
%%%%%%%%%%%%%%%%%%%%%%%%%%%%%%%
\section{First proof}
We first review the general result
for the Hubbard model on a line
graph proved in \cite{Andreas}.

We start from abstract notations.
Let $G=(V,E)$ be an abstract graph,
where $V$ is the set of vertices (sites)
 $\alpha,\beta,\ldots\in V$, and
$E$ is the set of edges (bonds) which are nothing but pairs of
vertices like $\{\alpha,\beta\}$.
Given a graph $G$, one can construct the
corresponding line graph $L(G)=(V_{\rm L},E_{\rm L})$
by the following procedure.
The set of vertices (sites) $V_{\rm L}$ (whose elements are denoted as
$x,y,\ldots\in V_{\rm L}$) is taken to be identical to the set $E$.
This means that we identify edges in $G$ with the vertices (sites) in
$L(G)$ as, for example, $x=\{\alpha,\beta\}$, $y=\{\alpha,\gamma\}$,
etc.
Next we declare that two vertices $x,y\in V_{\rm L}$ are adjacent to
each other if the corresponding two edges in $E$ share a common vertex
in $V$.
The $x$ and $y$ in the above example are adjacent to each other since
the corresponding edges in $E$ have a common vertex $\alpha$.
$E_{\rm L}$ is the set of edges (bonds) in $L(G)$, which
consists of all the adjacent pairs (like $\{x,y\}$)
of vertices (sites) in $V_{\rm L}$.
Finally we set $M(G)=|E|-|V|+1$ if $G$ is bipartite\footnote{
$G$ is bipartite if it can be decomposed into two disjoint sublattices
as $G=A\cup B$ with the property that any edge in $E$ joins a vertex
in $A$ with a vertex in $B$.
}, and $M(G)=|E|-|V|$ if $G$ is non-bipartite.

We define the Hubbard model on the line graph $L(G)$.
With each site $x\in V_{\rm L}$, we associate fermion operator
$c_{x,\sigma}$, and consider the Hamiltonian
\begin{equation}
	H = t \sum_{\{x,y\}\in E_{\rm L}}\sum_{\sigma=\uparrow,\downarrow}
	c^{\dagg}_{x,\sigma}c_{y,\sigma}
	+
	U\sum_{x\in V_{\rm L}}n_{x,\uparrow}n_{x,\downarrow}.
	\label{Ham2}
\end{equation}
Then the main result of \cite{Andreas} is the following.
\par\bigskip\noindent
{\bf Theorem 1:}
Suppose that the graph $G$ is twofold connected\footnote{
A graph is twofold connected if and only if one cannot make it
disconnected by a removal of a single vertex.
}.
Then in the Hilbert space where the electron number is fixed to
$N_{\rm e}=M(G)$, the ground states of (\ref{Ham2}) exhibit
saturated ferromagnetism
and are nondegenerate apart from the trivial spin degeneracy for any
$t>0$ and $U>0$.
\par\bigskip
This theorem applies to the Hubbard model defined on various line
graphs, a typical one being the kagom\'{e} lattice.

By construction, a general line graph is a graph that consists
of complete graphs connected at the vertices such that
two complete graphs have some vertices in common.
Thus the complete graph is the most trivial line graph.
It can be constructed taking
graph $G$ which has only two vertices, and has $N$
edges joining them.
More precisely we
set $V=\{\alpha,\beta\}$, and $E$ to be the set consisting of $N$
identical copies of the edge $\{\alpha,\beta\}$.
The corresponding $V_{\rm L}$ consists of $N$ sites, and any pair of
sites are adjacent with each other.
Since $G$ is bipartite, we get $M(G)=|E|-|V|+1=N-2+1=N-1$.
Thus Theorem 1 precisely reduces to the Corollary in Section~\ref{pr}.

The proof of the Corollary in \cite{Pieri}
is based on a theorem in \cite{am2},
which is nothing but a generalization of the above Theorem 1.
The structure of the proof in \cite{am2} is the same as
in \cite{Andreas}.
%%%%%%%%%%%%%%%%%%%%%%%%%%%%%%%
\section{Second proof}
In \cite{AndreasHal}, a class of Hubbard models with
ferromagnetism which is
stable against the change of electron density was
discussed.
It was noted in the Remark on pages 355 and 356 that ferromagnetism
can be established for more general
models defined on lattices with certain cell structure\footnote{
The proof was omitted in \cite{AndreasHal} since it was a
straightforward extension of the one for the models mainly discussed
in \cite{AndreasHal}.
The proof was supplied for completeness in Section 5.4 of \cite{Hal}.
}.

A trivial (and the least interesting)
version of this model is the one consisting of a single cell.
Let us reproduce it here (in a slightly generalized form)
for the readers' convenience.
Let $\Lambda=\{1,2,\ldots,N\}$ be the set of $N$ sites.
For $i=1,2,\ldots,N$, we let $\lambda_{i}$ be an arbitrary
nonvanishing (complex) quantity.
We take the Hamiltonian
\begin{equation}
	H
	=
	t \sum_{\sigma=\uparrow,\downarrow}
	\left(\sum_{i\in\Lambda}\lambda_{i}c_{i,\sigma}\right)^{\dagg}
	\left(\sum_{i\in\Lambda}\lambda_{i}c_{i,\sigma}\right)
	+U \sum_{i\in\Lambda}n_{i,\uparrow}n_{i,\downarrow}.
	\label{Ham3}
\end{equation}
Then we have\footnote{
In the proof, we regard an arbitrary site as the internal site, and
the rest as the external sites.
}
\par\bigskip\noindent
{\bf Theorem 3:}
In the Hilbert space where  the electron number is fixed to
$N_{\rm e}=N-1$, the ground states of (\ref{Ham3}) exhibit
saturated ferromagnetism
and are nondegenerate apart from the trivial spin degeneracy for any
$t>0$ and $U>0$.
\par\bigskip
This reduces to the Corollary in Section~\ref{pr}
if we set $\lambda_{i}=1$
for all $i$.
%%%%%%%%%%%%%%%%%%%%%%%%%%%%%%%
\section{Third proof}
There is another
straightforward proof that makes use of the wellknown
Nagaoka's theorem \cite{Nagaoka}.
The theorem (in its most general form)
applies to the $U\to\infty$
limit of the Hamiltonian (\ref{Ham}), and states that the ground
states exhibit ferromagnetism and are nondegenerate apart from the
trivial spin degeneracy.

One can easily check that the state
\begin{equation}
	\Phi_{\uparrow}
	=
	\left(\prod_{i=1}^{N-1}
	(c^{\dagg}_{i,\uparrow}-c^{\dagg}_{N,\uparrow})
	\right)
	\Phi_{\rm vac}
	\label{Phiup}
\end{equation}
minimizes both the hopping part and the interaction part
of the Hamiltonian (\ref{Ham}) when $t>0$.
Therefore $\Phi_{\uparrow}$ is a ground state of (\ref{Ham}) for any
$U\ge0$, and the only nontrivial task in the proof of
the Corollary in Section~\ref{pr} is to
show that there are no other ground states.
Given Nagaoka's theorem, this is easy.

Assume that, for some $U>0$,
 there is a ground state which is not an SU(2) rotation of
$\Phi_{\uparrow}$.
Since the ground state energy does not depend on $U$, it must remain
to be ground state for any larger $U$.
But this contradicts to the Nagaoka's theorem.
%%%%%%%%%%%%%%%%%%%%%%%%%%%%%%%

%%%%%%%%%%%%%%%%%%%%%%%%%%%%%%%

\begin{thebibliography}{1}
\bibitem{Salerno}
M. Salerno, Ferromagnetic ground states of the Hubbard model
on a complete graph, preprint, solv-int/9603008 (1996)
\bibitem{Wang}
D.F. Wang, Ferromagnetism in an itinerant electron system: Hubbard model
on complete graph, preprint, cond-mat/9604121 (1996)
\bibitem{Pieri}
P. Pieri, Ferromagnetism in the Hubbard model with an
infinite range hopping, preprint, cond-mat/9605156 (1996)
\bibitem{Andreas}
A. Mielke,
Ferromagnetism in the Hubbard model on line graphs and further
considerations,
J. Phys. A: Math. Gen. {\bf 24}, 3311--3321 (1991).
\bibitem{am2}
A. Mielke, Ferromagnetism in the Hubbard model and Hund's rule,
Phys. Lett. A {\bf 174}, 443--448 (1993)
\bibitem{AndreasHal}
A. Mielke and H. Tasaki,
Ferromagnetism in the Hubbard model:
Examples from models with degenerate single-electron ground states,
Commun. Math. Phys. {\bf 158}, 341--371 (1993).
\bibitem{Nagaoka}
Y. Nagaoka,
Ferromagnetism in a narrow almost half--filled s band,
Phys. Rev. {\bf 147}, 392--405 (1966).
For a general form see H. Tasaki,
Extension of Nagaoka's theorem on the large--$U$ Hubbard model,
Phys. Rev. {\bf B40}, 9192--9193 (1989)
\bibitem{Hal}
H. Tasaki,
Stability of ferromagnetism in Hubbard models with nearly-flat bands,
J. Stat. Phys. {\bf 84}, nos. 3/4 (1996), in press.
\bibitem{later}
H. Tasaki,
The Hubbard model: Introduction and some rigorous results,
preprint, cond-mat/9512169.
\end{thebibliography}
\end{document}